\documentclass[preprint, floatfix, showpacs]{revtex4}

\usepackage[final]{epsfig}
\usepackage{amsmath}
\usepackage{parskip}

\begin{document}

\title{Photonic measurements of the longitudinal expansion dynamics in Heavy-Ion collisions}

\author{Thorsten Renk}

\preprint{DUKE-TH-04-276}

\affiliation{Department of Physics, Duke University, PO Box 90305,  Durham, NC 27708 , USA}

\begin{abstract}
Due to the smallness of the electromagnetic coupling, photons escape from the hot and dense
matter created in an heavy-ion collision at all times, in contrast to hadrons which
are predominantly emitted in the final freeze-out phase of the evolving system.
Thus, the thermal photon yield carries an imprint from the early evolution. We suggest
how this fact can be used to gain information about where
between the two limiting cases of Bjorken (boost-invariant expansion) and
Landau (complete initial stopping and re-expansion) hydrodynamics the actual evolution 
can be found. We argue that both the rapidity dependence of the photon yield and 
photonic HBT radii are capable of answering this question. 
\end{abstract}

\maketitle

\section{Introduction}
\label{sec_introduction}

The boost-invariant hydrodynamic model proposed by Bj{\o}rken \cite{Bjorken} for the
description of ultrarelativistic heavy-ion collisions is frequently used at
RHIC energies for
estimates of the initial energy density in heavy-ion collisions or the
lifetime from the measured Hanbury-Brown Twiss (HBT) correlation radius 
$R_{long}$ \cite{Sinyukov} as well as in hydrodynamical descriptions
of the evolving system (see. e.g. \cite{Dinesh}).

While the original notion of boost-invariance is an asymptotic concept,
its application to RHIC energies usually implies two things: 1) the 
distribution of matter in some finite interval around midrapidity is assumed
to be (almost) independent of rapidity and 2) the longitudinal dynamics is assumed to be
unaccelerated expansion which in turn means that momentum rapidity
$y = \frac{1}{2}\ln \frac{E+p_z}{E-p_z}$ is always equal spacetime
rapidity $\eta_s = \frac{1}{2}\ln \frac{t+z}{t-z}$ (this is not
true in the presence of longitudinal acceleration).

In contrast, charged meson rapidity distributions as obtained by the
BRAHMS collaboration \cite{Brahms} do not show a flat plateau around
midrapidity  even at top RHIC energy. The
distributions are however well described by Landau hydrodynamics \cite{Landau}
as argued in \cite{Steinberg}. Likewise there is no boost invariance
seen in the rapidity dependence of elliptic flow as measured by the
PHOBOS collaboration \cite{Phobos}. 

In a model framework adjusted to reproduce the full set of observables
characterizing the hadronic freeze-out, i.e. single particle transverse
mass spectra and rapidity distributions and two particle HBT correlation radii
\cite{Renk} it was found that simultaneous agreement with all data sets can
only be achieved if the assumption of a boost-invariant expansion is
dropped. In fact, a sizeable difference of $\Delta y = 2\cdot 1.8$
between initial and final width of the source in momentum space rapidity
is required.

This, however, is rather indirect evidence since it rests on a backward 
extrapolation of the observed final state. In contrast, thermal photons
would offer the opportunity to test the longitudinal evolution directly \cite{Rischke}.
The essential idea is as follows: In a Landau scenario, the source
is initially  very narrow around midrapidity. Since the hard photon
emission rate is strongly temperature dependent, the dominant contribution
to the photon yield arises from early times. Thus, we expect that the
hard photon yield as a function of rapidity shows a thermal smearing
of the initial (narrow) source extension in rapidity. In contrast, in
a boost-invariant expansion we expect a much broader distribution
reflecting the initial distribution of matter across a large rapidity
interval.

There is an additional factor which needs to be taken into account: Due
to its large initial extension in rapidity, a Bj{\o}rken scenario leads
to much more rapid cooling than a Landau one. Hence, while in a Landau
scenario the hot, early phase will be dominant, this is not so in a Bj{\o}rken
framework. The different weights of the contributions of early times and
late times are expected to leave a characteristic imprint on HBT 
correlation radii measured even at midrapidity.

In this work, we discuss both ideas and demonstrate what predictions
for the photonic observables can be made using either the scenario
determined from a fit to spectra and HBT in \cite{Renk} or a Bj{\o}rken or
a Landau one.

\section{The model framework}

Several calculations studying photon emission based on a hydrodynamical
fireball evolution model have been made so far for different collision systems
and energies\cite{Hydro, Hydro1, Hydro2, Hydro3}. In the present study investigating 200 AGeV AuAu collisions, 
we will use a parametrized evolution model instead which allows for a complete
description of hadronic transverse mass spectra as well as HBT correlation
parameters \cite{Renk} and which can easily be tuned to interpolate between
Bj{\o}rken and Landau dynamics.

The model for the evolution of hot matter is 
described in detail in \cite{Renk, Synopsis}. Here we only present the
essential outline and focus on (almost) central collisions:

For the entropy density at a
given proper time we make the ansatz 
\begin{equation}
s(\tau, \eta_s, r) = N R(r,\tau) \cdot H(\eta_s, \tau)
\end{equation}
with $\tau $ the proper time measured in a frame co-moving with a given volume element 
 and $R(r, \tau), H(\eta_s, \tau)$ two functions describing the shape of the distribution
and $N$ a normalization factor. We use Woods-Saxon distributions 
\begin{equation}
\begin{split}
&R(r, \tau) = 1/\left(1 + \exp\left[\frac{r - R_c(\tau)}{d_{\text{ws}}}\right]\right)
\\ & 
H(\eta_s, \tau) = 1/\left(1 + \exp\left[\frac{\eta_s - H_c(\tau)}{\eta_{\text{ws}}}\right]\right)
\end{split}
\end{equation}
for the shapes. Thus, the ingredients of the model are the skin thickness 
parameters $d_{\text{ws}}$ and $\eta_{\text{ws}}$
and the para\-me\-tri\-zations of the expansion of 
the spatial extensions $R_c(\tau), H_c(\tau)$ 
as a function of proper time. From the distribution of entropy density, the thermodynamics can be inferred
via the EoS and particle emission is then calculated using the Cooper-Frye formula.
For simplicity, we 
assume that the flow is built up by a constant acceleration $a_\perp$, hence 
$R_c(\tau) = R_c^0 + \frac{a_\perp}{2} \tau^2$
with $R_c^0$ an initial radial extension as found in overlap calculations.
The rapidity distribution is assumed to grow from some initial width $2 \cdot y_0$
to a final width $2 \cdot y_F$. This determines the extension of the emitting source in spacetime
rapidity $\eta_s$ \cite{Renk, Synopsis}.

In \cite{Renk}, the model parameters have been adjusted such that the model gives a good
description of the data. This implies an initial rapidity width of $y_0 = 1.7$.
 In order to compute a Bj{\o}rken scenario, we set the
initial width of the rapidity distribution equal to the final distribution width
$y_0 = y_F$. For a Landau scenario we choose $y_0 = 0$. In both cases
we readjust the model parameters such that the single particle spectra
are reproduced (this implies  losing agreement with the HBT data).

The spectrum of emitted photons can be found by folding the photon emission rates
for the quark-gluon plasma (QGP) phase \cite{Moore} and for a hot hadronic gas \cite{Gale} 
with the
fireball evolution. In order to account for flow, the energy of a photon emitted
with momentum $k^\mu =(k_t, {\bf k_t}, 0)$ has to be evaluated in the local rest
frame of matter, giving rise to a product $k^\mu u_\mu$ with $u_\mu(\eta_s, r, \tau)$
the local flow profile. Following the results in \cite{Renk} we assume 
for the spatial dependence of the flow field the relations
$y = f(\tau) \cdot \eta_s$ and $y_\perp = g(\tau) \cdot r$ with 
$y_\perp$ the transverse
rapidity and $f,g$ two functions determined by the evolution. 
The distribution of entropy density is manifest
in the dependence of the temperature $T = T(\eta_s, r, \tau)$ on the spacetime
position. In order to account for the breakup of the system once a temperature
$T_F$ is reached, a factor $\theta(T-T_F)$ has to be included into the folding
integral. 

Using the folding integral of the rate with the fireball evolution
as emission function $S(x,K)$ (describing the amount of photons
with momentum $K^\mu$ emitted at spacetime point $x^\mu$) 
we calculate the HBT parameters as \cite{HBTReport, HBTBoris}
\begin{equation}
R_{\text{side}}^2 = \langle \tilde{y}^2 \rangle \quad R_{\text{out}}^2 =\langle (\tilde{x} -\beta_\perp \tilde{t})^2 
\rangle \quad R_{\text{long}} = \langle \tilde{z}^2 \rangle
\end{equation}
with $\tilde{x}_\mu = x_\mu -\langle x_\mu \rangle$ and
\begin{equation}
\langle f(x)\rangle(K) = \frac{\int d^4 x f(x) S(x,K)}{\int d^4x S(x,K)}.
\end{equation}

\section{Rapidity dependence of hard thermal photon emission}

\begin{figure*}[htb]
\epsfig{file=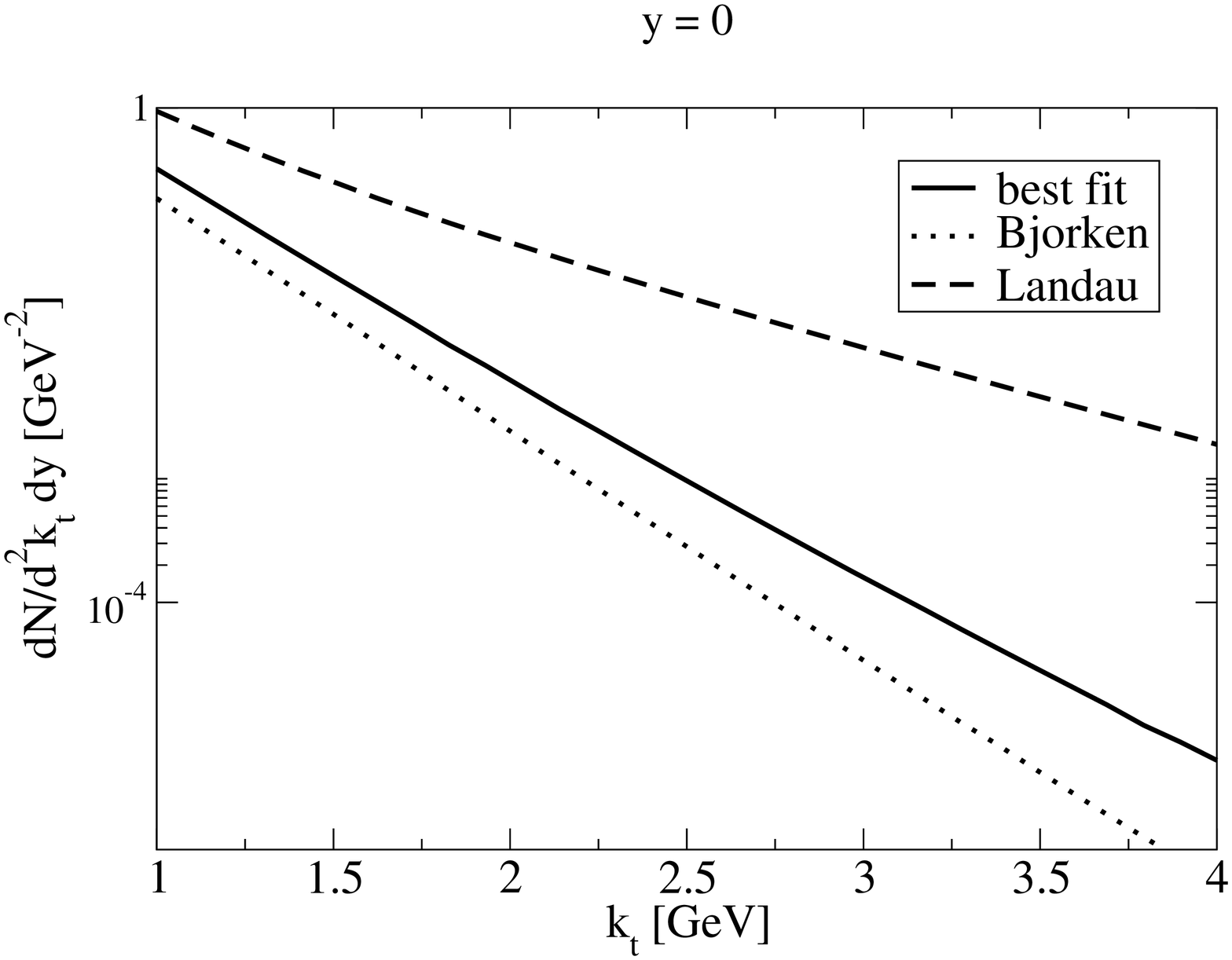, width=7.8cm}
\epsfig{file=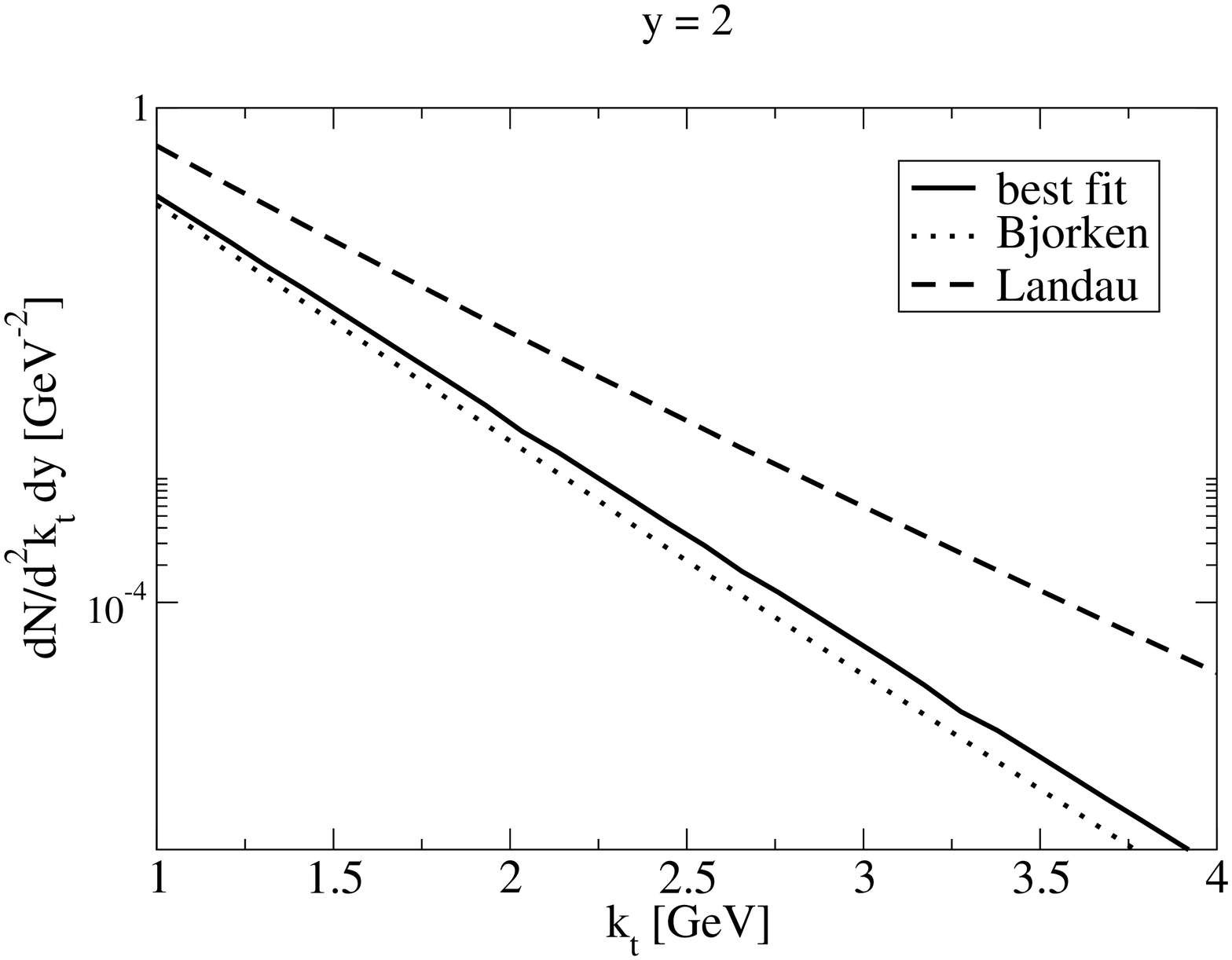, width=7.8cm}
\caption{\label{F-1}The hard thermal photon spectrum at midrapidity (y=0. left panel) and
forward rapidity (y=2, right panel) for the best fit scenario described in \cite{Renk}, 
a Bj{\o}rken and a Landau scenario.}
\end{figure*}

We show the resulting spectra of hard thermal photons in the momentum range
between 1 and 4 GeV in Fig.~\ref{F-1} at two different rapidities.

It is instructive to observe that both slope and absolute yield changes
strongly as a function of $y$ for the Landau scenario. This reflects the fact that the initial
high temperature phase (leading to a relatively flat slope) never
radiates out into the $y=2$ slice --- only in the later stages when
hot matter expands across $y=2$ there is a significant
contribution, albeit from matter with a much lower temperature, leading to
a steeper spectral slope and reduced yield.

In contrast, the photon yield from a Bj{\o}rken scenario is practically
unchanged as a function of rapidity, reflecting the approximate 
boost-invariance.

\begin{figure}[htb]
\epsfig{file=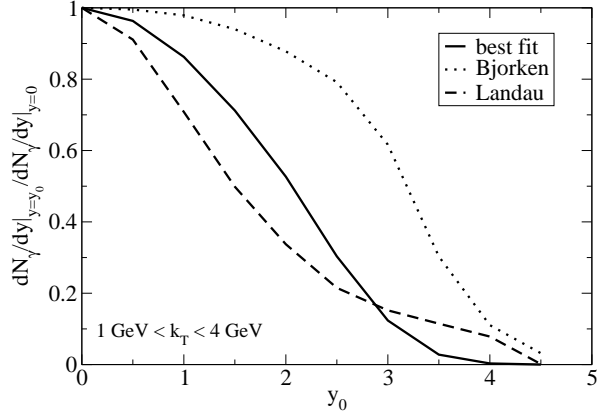, width=7.8cm}
\caption{\label{F-2}Integrated yield (1.0 GeV $< k_T <$ 4 GeV) 
of thermal photons as a function of rapidity $y_0$.}
\end{figure}

In order to highlight the differences more clearly we show in
Fig.~\ref{F-2} the $k_T$-integrated yield (1 GeV $< k_T <$ 4 GeV)
at rapidity $y_0$ divided by the integrated yield at midrapidity.
This choice has the additional advantage that model-dependences such as the
precise normalization of the emission rates tend to cancel out.

The different longitudinal source structure is now directly apparent.
The Landau scenario is characterized by thermal smearing of about 1 unit
of rapidity of a source at midrapidity (without any longitudinal
flow) whereas the Bj{\o}rken scenario shows the broad distribution
of matter across $\sim 3$ units of rapidity at all times. A
measurement of the thermal photon yield at midrapidity and at 
$y_0 = 2$ would well
be capable of making a distinction between the three scenarios.

\section{Hard thermal photon HBT at midrapidity}

HBT correlation measurements do not measure the true geometrical size
of the source but rather a region of homogeneity \cite{HBTReport, HBTBoris}
which is only identical with the geometry for vanishing flow
gradients in the source. For finite flow gradients, the measured
correlation radii show a characteristic falloff with the correlated pair
momentum $k_T$. The precise shape of the correlation radii as a function
of transverse momentum results from a complex interplay between temperature
and flow during the whole evolution.

Nevertheless, we can formulate some basic expectations. Due to the
high initial compression, the hard photon yield from a Landau scenario
is expected to be dominated by the initial phase of the expansion.
In this phase, however, there is no significant transverse flow (which
builds up gradually driven by transverse pressure) and the geometrical
size of the source in longitudinal direction is very small 
(for complete stopping it is given by
the Lorentz-contracted size of the overlapping nuclei). Thus, we 
would expect only a weak falloff of $R_{side}$ with $k_T$ 
and $R_{long}$ to be determined primarily by the spatial resolution scale
of photons with a given momentum.

In contrast, a Bj{\o}rken expansion may well receive significant relative
contributions to the yield from later stages
due to the shorter duration of the inital hot phase.
This would imply a slightly
larger $R_{side}$ for vanishing $k_T$ but a stronger falloff with
$k_T$ and an increased value $R_{long}$ as compared to the initial
size at equilibration time. The relevant underlying scale 
for $R_{side}$ is in all cases the nuclear overlap radius.

\begin{figure*}[htb]
\epsfig{file=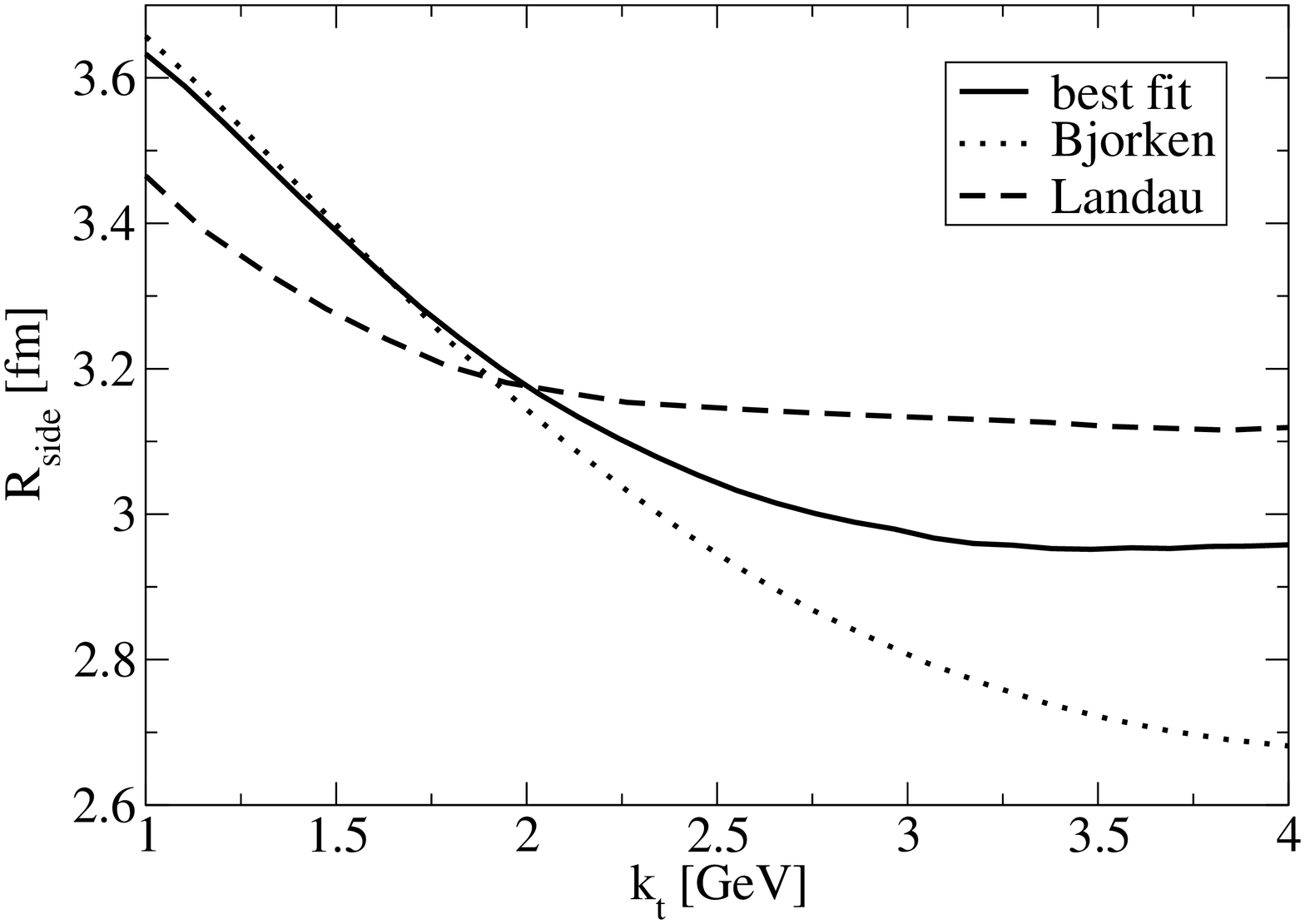, width=7.8cm}
\epsfig{file=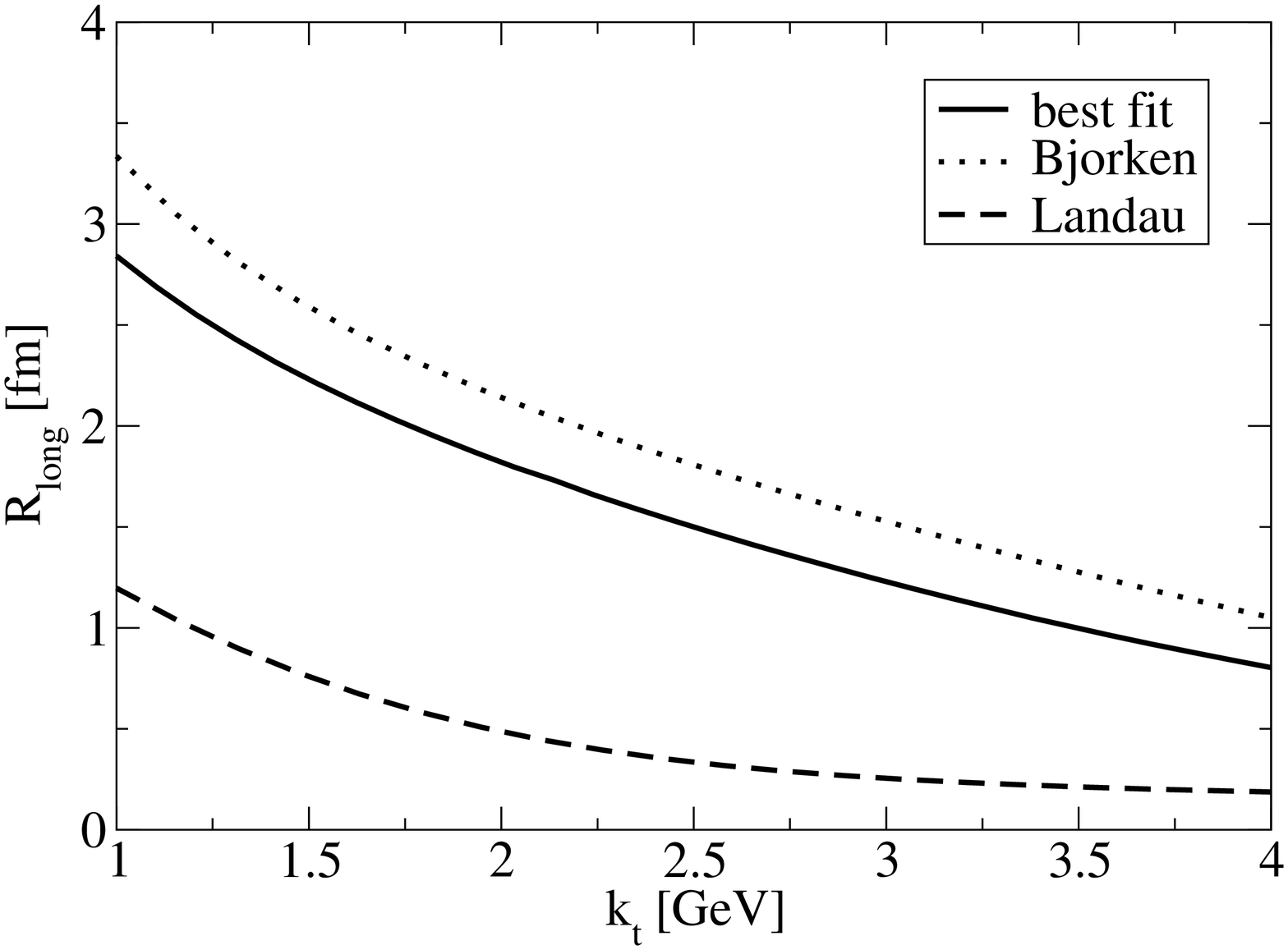, width=7.8cm}
\caption{\label{F-3}Hard photon HBT correlation radii $R_{side}$ (left panel)
and $R_{long}$ (right panel) for the best fit scenario described 
in \cite{Renk}, a Bj{\o}rken and a Landau scenario.}
\end{figure*}

The result of the calculation can be seen in Fig.~\ref{F-3}. To a good degree,
the expected behaviour is indeed seen. In particular, the different
falloff of $R_{side}$ for $k_T > 2.5$ GeV appears to be a good indicator
of the longitudinal dynamics. $R_{long}$ in contrast is presumably only capable
of identifying a scenario very similar to a Landau one, otherwise the
qualititive behaviour of the different curves is too similar. Note that the
observed $R_{long}$ for the Landau scenario could not be as small as shown in
the plot due to constraints posed by the uncertainty relation which doesn't allow
to narrow down the photon emission region to arbitrary small size.

\section{The role of pre-equilibrium photons}

It is well known that in addition to thermal photons a sizeable
contribution of prompt photons (calculable in perturbative QCD)
is expected to contribute to the hard photon yield, and various
attempts have been made to calculate its magnitude (see e.g. 
\cite{Dumitru, Wong, Srivastava}). This contribution might
well outshine the signals proposed here and change the
conclusions.

In order to address this question carefully, we do not only have to take
into account the primary hard scattering processes as a potential source
of photons but also hard re-scattering processes as the system approaches
equilibrium. Therefore we use here the VNI/BMS parton cascade
model (PCM) to estimate the role of pre-equilibrium hard photon production 
\cite{VNI-Photons, VNI-HBT}.

There is still the caveat that the re-scattering described by the PCM does not
lead to a Landau-like stopping of the incoming matter, nevertheless
we use the results to gain some intuition in the orders of magnitude
involved.

Including the LPM suppression in the PCM, we find that thermal photons
may dominate the yield below 2--2.5 GeV for the best fit and the Bj{\o}rken
scenario whereas they would dominate the yield in the whole
momentum range for a pure Landau evolution \cite{VNI-NEW}.

Since the photon yield drops (almost) exponentially with $k_t$, this
implies that the rapidity dependence of the integrated yield would
still be a reliable signal (being dominated by the low $k_T$ yield).

However, the behaviour of the HBT correlation radii in the
interesting region above 2 GeV is likely to be distorted
by pre-equilibrium photons (which would incidentially resemble
Landau dynamics as they are characterized by small transverse
flow).

Turning the argument around, a simultaneous measurement
of the HBT correlations at midrapidity and of the integrated yield at forward
rapidity could still provide valuable insight into the magnitude of
the pre-equilibrium contribution at different momenta. 
A detailed investigation of this question
is however beyond the scope of this work.

\section{Summary}

We have argued that photons provide a direct measurement of the early
longitudinal dynamics of a heavy-ion collision which can otherwise
only be inferred indirectly from hadronic probes. The underlying reason
is that due to the smallness of the electromagnetic coupling the
measured photon yield represents an integral over the whole
fireball evolution rather than a snapshot at breakup.

In particular, we have argued that the rapidity dependence of the hard
photon yield is a good probe to distinguish between Landau and Bj{\o}rken-like
dynamics since it directly reveals the rapidity extension of the
emission source. Since we compare the rapidity dependence of a ratio of integrated
yields many uncertainties associated with the calculation of emission rates
drop out and the result mainly reflects kinematic properties of the source.

In addition, we have investigated the potential of using HBT correlation
measurements at midrapidity to investigate the longitudinal
evolution. HBT correlations show what part of the evolution
dominates the photon yield rather than directly reflecting longitudinal 
dynamics. We found that the falloff of $R_{side}$ with $k_t$ above
2.5 GeV would indeed give a good indication if the photon emission
is dominated by matter without significant transverse flow or not,
however this signal is easily obscured by pre-equilibrium
photon emission which would never show significant transverse
flow.

Nevertheless, both measuring the rapidity dependence of the hard photon yield
and the HBT correlation parameters at midrapidity are capable of
revealing interesting details of the early fireball evolution which
cannot easily be studied otherwise.


\begin{acknowledgments}

I would like to thank C.~Gale, S.~Turbide, J.~Ruppert, 
S.~A.~Bass and B.~M\"{u}ller for helpful 
discussions and comments  during the preparation of this paper. 
This work was supported by the DOE grant DE-FG02-96ER40945 and a Feodor
Lynen Fellowship of the Alexander von Humboldt Foundation.
\end{acknowledgments}

\end{document}